\documentclass[12pt]{article}

\usepackage{amssymb}
\usepackage[dvips]{graphicx}

\newcommand {\beq}{\begin{equation}}
\newcommand {\eeq}{\end{equation}}
\newcommand {\beqa}{\begin{eqnarray}}
\newcommand {\eeqa}{\end{eqnarray}}
\newcommand {\n}{\nonumber \\}
\newcommand {\Real}{\mbox{Re}}
\newcommand {\Imag}{\mbox{Im}}
\newcommand {\Tr}{\mbox{Tr}}
\newcommand {\tr}{\mbox{tr}}
\newcommand {\Pf}{\mbox{Pf}}

\newcommand {\ee}{\mbox{e}}

\newcommand {\dd}{\mbox{d}}

\newcommand {\del}{\partial}

\renewcommand{\theequation}{\thesection.\arabic{equation}}

\begin{document}
\setlength{\oddsidemargin}{0cm}
\setlength{\baselineskip}{7mm}

\begin{titlepage}
 \renewcommand{\thefootnote}{\fnsymbol{footnote}}
\begin{normalsize}
\begin{flushright}
\begin{tabular}{l}
NBI-HE-00-24\\
NORDITA-2000/50-HE\\
hep-th/0005147\\
May 2000
\end{tabular}
\end{flushright}
  \end{normalsize}

\vspace*{0cm}
    \begin{Large}
       \begin{center}
         {Monte Carlo Studies of the IIB Matrix Model at Large $N$}
       \end{center}
    \end{Large}
\vspace{2mm}

\begin{center}
J. A{\scshape mbj\o rn}$^{1)}$\footnote
            {e-mail address : ambjorn@nbi.dk},
K.N. A{\scshape nagnostopoulos}$^{2)}$\footnote
            {e-mail address : konstant@kiritsis.physics.uoc.gr},\\
W. B{\scshape ietenholz}$^{3)}$\footnote
            {e-mail address : bietenho@nordita.dk},
T. H{\scshape otta}$^{4)}$\footnote
            {e-mail address : hotta@hep1.c.u-tokyo.ac.jp}
           {\scshape and}
           J. N{\scshape ishimura}$^{1)}$\footnote{
Permanent address : Department of Physics, Nagoya University,
Nagoya 464-8602, Japan,\\
e-mail address : nisimura@nbi.dk}\\
      \vspace{1cm}
        $^{1)}$ {\itshape Niels Bohr Institute, Copenhagen University,} \\
              {\itshape Blegdamsvej 17, DK-2100 Copenhagen \O, Denmark}\\
        $^{2)}$ {\itshape Department of Physics, University of Crete,}\\
              {\itshape P.O. Box 2208, GR-71003 Heraklion, Greece}\\
        $^{3)}$ {\itshape NORDITA} \\
              {\itshape Blegdamsvej 17, DK-2100 Copenhagen \O, Denmark} \\
        $^{4)}$ {\it Institute of Physics, University of Tokyo,}\\
                 {\it Komaba, Meguro-ku, Tokyo 153-8902, Japan}
\end{center}

\hspace{5cm}


\begin{abstract}
\noindent
%
%
The low-energy effective theory of the IIB matrix model
developed by H.~Aoki et al. is written down explicitly 
in terms of bosonic variables only.
The effective theory is then studied
by Monte Carlo simulations in order
to investigate
the possibility of a
spontaneous breakdown of Lorentz invariance.
The imaginary part of the effective action,
which causes the so-called sign problem in the simulation, 
is dropped by hand.
%
%
The extent of the eigenvalue distribution of the bosonic matrices
shows a power-law large $N$ behavior, consistent with a simple
branched-polymer prediction.
We observe, however, that the eigenvalue distribution becomes
more and more isotropic in the ten-dimensional space-time 
as we increase $N$.
This 
suggests 
that if the spontaneous breakdown of Lorentz invariance
really occurs in the IIB matrix model, 
a crucial r\^{o}le must be played
by the imaginary part of the effective action.
\end{abstract}
\vfill
\end{titlepage}
\vfil\eject

\setcounter{footnote}{0}
\section{Introduction}
\setcounter{equation}{0}

\renewcommand{\thefootnote}{\arabic{footnote}} 

For more than two decades 
superstring theories have been studied as the most promising candidates
for a unified theory of all the interactions including gravity.
The theories have the potential to 
predict
the space-time dimensionality, the gauge group,
the matter content,
and so on, from first principles.
%
%
The existence of infinitely many perturbative vacua
implies, however, 
that an understanding of nonperturbative effects is crucial
to extract information about the real vacuum of the theory.
Recent proposals for nonperturbative formulations
of superstring 
theories \cite{BFSS,IKKT,nonperturbative,nonperturbative2,variants}
may therefore be of analogous importance for our understanding of 
non-perturbative aspects of string theory as 
lattice gauge theory \cite{Wilson} has been 
in understanding nonperturbative dynamics of gauge theories.
The IIB matrix model \cite{IKKT},
which is conjectured to be a nonperturbative formulation of
type IIB superstring theory (for a review, see Ref.~\cite{IIBrev}),
takes the form of a large $N$ reduced model \cite{EK},
and in the same way that Monte Carlo studies of lattice gauge theory
clarified many important nonperturbative aspects of
the strong interaction,
Monte Carlo studies of the IIB matrix model
might illuminate the nonperturbative dynamics of 
superstring theories\footnote{Monte Carlo studies of 
Matrix Theory \cite{BFSS} would be technically more involved
due to a lattice discretization of the time direction \cite{JW}.}.
A number of numerical studies have already been done 
\cite{SU-N,2DEK,KNS,KS,HNT,Eigen,OY,Potsdam,AABHN,KS2}
to pursue that direction. Earlier numerical studies 
of {\em world-sheet perturbative} aspects of superstring-like theories
can be found in Refs.~\cite{extrinsic,bowick,super}. 
%

In this paper, we study 
the IIB matrix model at large $N$
by Monte Carlo simulations.
In particular, we investigate the possibility of a spontaneous symmetry
breakdown of Lorentz invariance using the low-energy effective theory 
of the IIB matrix model proposed by
Ref.~\cite{AIKKT}.  
The authors of \cite{AIKKT} also study this issue, using an 
approach where the bosonic and fermionic matrices are both decomposed
into diagonal elements and off-diagonal elements,
and the off-diagonal elements are integrated out first perturbatively.
Such a perturbative expansion is valid when 
the bosonic diagonal elements, which can be regarded as 
coordinates of $N$ points in ten-dimensional flat space-time,
are well separated from each other.
In other words, what one obtains after integrating over
the off-diagonal elements perturbatively can be considered as
a low-energy effective theory of the IIB matrix model.
Note
that this perturbative expansion
has nothing to do with the string perturbative expansion
with respect to the world-sheet topology.
Therefore, even at the one-loop level,
the low-energy effective theory 
is expected to include
{\em nonperturbative} effects of superstring theory,
provided, of course, that the IIB matrix model conjecture is true.

In fact, in order to obtain the low-energy effective action only for the
bosonic diagonal elements,
one still has to perform the integration
over the fermionic diagonal elements, which is
nontrivial,
since their action turns out to be quartic.
Although the explicit form of the final low-energy effective theory
has not been derived,
the theory was shown \cite{AIKKT}
to be described by
some complicated branched-polymer like system,
typically involving a ``double-tree'' structure,
in a flat ten-dimensional space time.
Thus, even at the one-loop level\footnote{The validity of
the one-loop approximation for studying the low-energy dynamics
of supersymmetric large $N$ reduced models has been
demonstrated in Ref.~\cite{AABHN} through the study of 
the four-dimensional version of the IIB matrix model.
We will come back to this point is Section \ref{lowenergy}.}, 
the low-energy effective theory
contains highly nontrivial dynamics.
It was further argued that the double-tree structure
of the one-loop low-energy effective theory 
might be responsible for a collapse of the
distribution of the bosonic diagonal elements.
The first Monte Carlo results of a branched-polymer system 
with a double-tree structure
was reported in Ref.~\cite{IIBrev}.
%
%

Here, we write down explicitly the low-energy effective theory of 
the IIB matrix model in terms of bosonic variables only.
%
%
Instead of integrating over both bosonic and fermionic 
off-diagonal elements first,
we leave the bosonic off-diagonal elements unintegrated.
The action for the fermionic diagonal elements is then
still quadratic and can be integrated explicitly, yielding a Pfaffian.
Integration over the bosonic off-diagonal elements 
as well as the bosonic diagonal elements 
can be done by Monte Carlo simulation.
In other words, the bosonic off-diagonal elements play the r\^{o}le
of auxiliary variables,
which enable us to simulate the complicated branched-polymer like system
describing the dynamics of the bosonic diagonal elements.
%

The Pfaffian induced by the integration over the
fermionic diagonal elements is generically complex.
In general, when the action of a theory has a non-zero imaginary part,
the number of configurations needed to extract any information
increases exponentially with the system size, except in a few
situations, where alternative sampling methods can be invented 
\cite{complex,sign}.
This notorious technical problem in Monte Carlo simulations
is known as the ``sign problem''.
In fact, the problem exists already in Monte Carlo simulations
of the original IIB matrix model
and it is inherited by the low-energy effective theory.
In the present work, we simply use the absolute value of
the Pfaffian in order to avoid the sign problem
and examine only the effect of the modulus of the Pfaffian.
Our results suggest that there is no 
spontaneous symmetry breakdown (SSB) of Lorentz invariance.
{}From this we conclude that if the SSB ever occurs in the IIB matrix model,
the phase of the Pfaffian must play a crucial r\^{o}le.
We also study the six-dimensional version of the IIB matrix model
for comparison.
The conclusion is the same,
but we find an intriguing difference in the finite $N$ effects.

The paper is organized as follows.
In Section \ref{model}, we describe the definition of the IIB matrix 
model and review some important properties relevant for this work.
In Section \ref{lowenergy},
we derive the low-energy effective theory of the IIB matrix model
and explain the model we investigate by Monte Carlo simulations.
In Section \ref{results},
we present our results for the distribution of the bosonic
diagonal elements.
In particular, we discuss the possibility of SSB of Lorentz invariance.
Section \ref{summary} is devoted to a summary and conclusions.
In Appendix A, we explain the details of the algorithm we use
for the Monte Carlo simulation.
In Appendix B, we present some systematic studies
for optimizing the parameters involved in the algorithm.

\vspace*{1cm}

\section{The IIB matrix model}
\setcounter{equation}{0}
\label{model}

The IIB matrix model \cite{IKKT} is formally a zero-volume limit of
ten-dimensional pure ${\cal N}=1$ supersymmetric Yang-Mills theory.
The action, therefore, is given by
\begin{eqnarray}
\label{action}
Z_{\rm IIB} &=& \int \dd A ~  \ee ^{-S_{\rm b}} Z_{\rm f} [A] ~~~;~~~
Z_{\rm f} [A] = \int \dd \psi ~ \ee ^{- S_{\rm f} } \ ,  \\
\label{action_bosonic}
S_{\rm b} &=& -\frac{1}{4 g^2}  \tr ([A_{\mu},A_{\nu}]^{2}) \  , \\
\label{action_fermionic}
S_{\rm f}  &=& - \frac{1}{2 g^2} \,
\tr \left( \psi _\alpha ( \tilde{\Gamma}_{\mu})_{\alpha\beta} 
[A_{\mu},\psi _\beta] \right)  \ .
\end{eqnarray}
$A_\mu$ ($\mu = 1,\cdots,10$) and 
$\psi_\alpha$ ($\alpha = 1,\cdots , 16$) are
$N \times N$ traceless Hermitian matrices,
which can be expanded in terms of the generators 
$t^a$ of SU($N$) as
\beq
(A_{\mu })_{ij} = \sum_{a=1}^{N^2-1} 
A_{\mu}^{ a} \, (t^{a})_{ij} ~~~~~;~~~~~  
(\psi_\alpha ) _{ij} = \sum_{a=1}^{N^2-1} 
\psi_{\alpha}^{ a} \, (t^{a})_{ij} \ , 
\eeq
where $A_\mu ^{a}$ is a real variable and 
$\psi_{\alpha}^{a}$ is a real Grassmann variable.
We assume that the generators $t ^a$ are
normalized as $\tr (t^a t^b) = \delta _{ab}$.
The measure $\dd \psi$ in (\ref{action}) is defined by
\beqa
\dd \psi  &=&  
\prod_{\alpha = 1 }^{16} \prod_{a = 1 }^{N^2 -1}  \dd \psi_{\alpha}^{ a}  \n
&=& \prod_{\alpha = 1 }^{16}
\left[ \prod_{i<j} \{2 \, \dd \Real (\psi_\alpha)_{ij} \dd 
\Imag (\psi_\alpha)_{ij} \}  \prod_{i=1} ^N \{ \dd (\psi_\alpha)_{ii} \}
\, \delta \left( \frac{1}{\sqrt{N}} 
\sum _{i=1} ^N (\psi_\alpha)_{ii} \right) \right]  \ ,
\label{measure_fermion}
\eeqa
and similarly for $\dd A$.
The model (\ref{action}) appears after a Wick rotation,
so that the metric has Euclidean signature.
The $16 \times 16$ matrices $\tilde{\Gamma} _\mu$ are defined
by
\beq
\tilde{\Gamma} _\mu = \, {\cal C} \,  \Gamma_{\mu} \ ,
\eeq
where $\Gamma_{\mu}$ are
ten-dimensional gamma matrices after Weyl projection,
and the unitary matrix ${\cal C}$ is a 
charge conjugation matrix satisfying
\beq
{\cal C} \, \Gamma _\mu \, {\cal C}^\dag
= (\Gamma _\mu)^\top ~~~~~;~~~~~
{\cal C} ^\top = {\cal C}  \ .
\label{chargeconj}
\eeq
Due to (\ref{chargeconj}),
the matrices $\tilde{\Gamma} _\mu$ are symmetric.

An explicit representation of the gamma matrices is given by
\beqa
&~&\Gamma _1 = i \, \sigma_2  \otimes \sigma_2  
              \otimes \sigma_2  \otimes \sigma_2 ~;~
\Gamma _2 = i \, \sigma_2  \otimes \sigma_2
              \otimes {\bf 1} \otimes \sigma_1 ~;~
\Gamma _3 = i \, \sigma_2  \otimes \sigma_2 
              \otimes {\bf 1} \otimes \sigma_3 ~;~\n
&~&\Gamma _4 = i \, \sigma_2  \otimes \sigma_1
              \otimes \sigma_2 \otimes {\bf 1} ~;~
\Gamma _5 = i \, \sigma_2  \otimes \sigma_3
             \otimes \sigma_2 \otimes {\bf 1} ~;~
\Gamma _6 = i \, \sigma_2  \otimes {\bf 1}
              \otimes \sigma_1 \otimes \sigma_2  ~;~ \n
&~&\Gamma _7 = i \, \sigma_2  \otimes {\bf 1}
             \otimes \sigma_3 \otimes \sigma_2 ~;~
\Gamma _8 = i \, \sigma_1  \otimes {\bf 1}
              \otimes {\bf 1} \otimes {\bf 1} ~;~
\Gamma _9 = i \, \sigma_3  \otimes {\bf 1}
              \otimes {\bf 1} \otimes {\bf 1} ~;~\n
&~&\Gamma _{10} = {\bf 1}  \otimes {\bf 1} \otimes {\bf 1} \otimes {\bf 1} \ ,
\label{Gamma}
\eeqa
for which the charge conjugation matrix ${\cal C}$ becomes a unit matrix
and therefore $\tilde{\Gamma} _\mu =  \Gamma _\mu$.

The model has a manifest ten-dimensional 
Lorentz invariance,
by which we actually mean an SO(10) invariance.
$A_{\mu}$ transforms as a 
vector and
$\psi _\alpha$ transforms as a Majorana-Weyl spinor.
The model is manifestly supersymmetric,
and also has a SU($N$) symmetry
\beq
\label{SU_N}
A_\mu  \mapsto  V A_\mu V^\dag  ~~~;~~~
\psi _\alpha \mapsto V \psi _\alpha V^\dag  \   ,
\eeq
where $V\in \mbox{SU}(N)$.
All these symmetries are inherited from the super Yang-Mills theory
before taking the zero-volume limit.
In particular, the SU($N$) symmetry 
(\ref{SU_N}) is a remnant of the local
gauge symmetry.

The fermion integral $Z_{\rm f} [A]$ in (\ref{action})
can be obtained explicitly as
\beq
Z_{\rm f} [A] 
= \Pf \, {\cal M}  \ ,
\label{fulldet}
\eeq
where
\beqa
{\cal M}_{a \alpha ,  b \beta} &=& 
-  i \, \frac{1}{g^2} \, f_{abc} 
( \tilde{\Gamma}_{\mu})_{\alpha\beta}   
A_\mu ^c    \ ,    \n
f_{abc} &=& -i \, \tr (t^a[t^b, t^c]) \ .
\eeqa
The real totally-antisymmetric tensor $f_{abc}$ gives 
the structure constants of SU($N$)
and the matrix ${\cal M}_{a \alpha ,  b \beta}$
is a $16 \,  (N^2-1) \times 16 \, (N^2-1)$ anti-symmetric matrix, regarding
each of $(a\alpha)$ and $(b\beta)$ as a single index.


The convergence of the integration over the bosonic matrices 
in (\ref{action}) is nontrivial since 
the integration domain for the Hermitian matrices is non-compact.
Even for finite $N$ there is a potential danger of divergence 
when the eigenvalues of $A_\mu$ become large. 
This issue has been addressed in Ref.~\cite{AIKKT}
using one-loop perturbative arguments
which pointed to the finiteness of 
the IIB matrix model given by (\ref{action}) 
for arbitrary $N$.
This conclusion is in agreement with
an exact result available for $N=2$ \cite{SuyamaTsuchiya}
and a numerical result obtained for $N=3$ \cite{KNS}.
Thus
it is conceivable
that the above conclusion, obtained by  
one-loop arguments, holds in general.

Since the model is well-defined without any cutoff, 
the parameter $g$, which is the only parameter of the 
model, can be absorbed by rescaling the variables,
\beqa
\label{rescaleA}
A_\mu &=& 
g^{1/2} X _\mu  \ ,\\
\psi_\alpha &
=& g^{3/4} \Psi _\alpha  \ .
\eeqa
Therefore, $g$ is a scale parameter rather than
a coupling constant, i.e.\
the $g$ dependence of physical quantities is completely 
determined on dimensional grounds\footnote{The scale parameter
$g$ should be tuned appropriately as one sends $N$ to infinity
so that all the correlation functions of Wilson loops
have a finite large $N$ limit.
Whether such a limit really exists or not is 
one of the important dynamical issues, which was addressed 
in Ref.~\cite{AABHN} for the four-dimensional version of 
the IIB matrix model.}. 
In what follows, we take $g=1$ without loss of generality.


For comparison, we also study the six-dimensional version of the
IIB matrix model.
In this regard we recall that 
pure ${\cal N}=1$ supersymmetric Yang-Mills theory
can be also defined in 3D, 4D and 6D, as well as in 10D.
Hence, by taking a zero-volume limit of these theories,
we arrive at supersymmetric large $N$ reduced models
which are $D=3,4,6$ versions of the IIB matrix model.
Using the one-loop argument mentioned above, one concludes that
the model is ill-defined for $D=3$, but
well-defined for $D=4,6,10$, irrespectively of $N$.
For $D=4$, Monte Carlo simulations up to $N=48$
further confirms this statement \cite{AABHN}.
The effective action induced by fermions 
(logarithm of the fermion integral $Z_{\rm f} [A]$)
is real for $D=4$.
It is complex in general for $D=6$ and $D=10$, however,
which causes the sign problem in Monte Carlo simulations.

\vspace*{1cm}

\section{Low-energy effective theory}
\setcounter{equation}{0}
\label{lowenergy}

In this section, we derive the low-energy effective theory of
the IIB matrix model (\ref{action})
along the lines discussed in Ref.~\cite{AIKKT}.
We first decompose the $N\times N$ Hermitian matrices 
$A_{\mu}$ and $\psi _\alpha$ as
\beqa
A_\mu  &=& \hat{x}_\mu  + \hat{a}_\mu  \ , \n
\psi _\alpha  &=& \hat{\xi}_\alpha  + \hat{\varphi}_\alpha \ ,
\label{docompose}
\eeqa
where $\hat{x}_\mu $ and $\hat{\xi}_\alpha$ 
represent the diagonal parts, while 
$\hat{a}_{\mu }$ and $\hat{\varphi}_{\alpha}$ 
represent the off-diagonal parts.
We also introduce 
$N$ ten-dimensional vectors $x_i$ ($i=1,\cdots , N$)
through $x_{i\mu} = (\hat{x}_\mu)_{ii}$,
and $N$ ten-dimensional Majorana-Weyl spinors 
$\xi_i$ ($i=1,\cdots , N$)
through $\xi_{i\alpha} = (\hat{\xi}_\alpha)_{ii}$.
For the off-diagonal elements, we use the notations
$a_{\mu ij}= (\hat{a}_\mu)_{ij} $ and
$ \varphi_{\alpha ij} = (\hat{\varphi}_\alpha)_{ij}$, where $i \neq j$.
Using the decomposition (\ref{docompose}),
the actions (\ref{action_bosonic}) and (\ref{action_fermionic}) 
can be written as
\beqa
S_{\rm b} &=&
 -  \tr \left(
-\frac{1}{2} \hat{a}_{\nu} [\hat{x}_{\mu},[\hat{x}_{\mu},\hat{a}_{\nu}]]
-\frac{1}{2} [\hat{x}_\mu, \hat{a}_\mu]^2
 +  [\hat{x}_{\mu},\hat{a}_{\nu}][\hat{a}_{\mu},\hat{a}_{\nu}] 
 + \frac{1}{4} [\hat{a}_{\mu},\hat{a}_{\nu}]^2   \right)  \ ,
\label{bosonicaction}
\\
S_{\rm f} &=&
- \frac{1}{2} \, (\tilde{\Gamma} _{\mu})_{\alpha\beta} \, 
\tr \Bigl( \hat{\varphi} _\alpha [\hat{x}_{\mu},\hat{\varphi} _\beta] 
 - \hat{\varphi} _\alpha  [\hat{\xi} _\beta ,\hat{a}_{\mu}] 
 - \hat{a}_{\mu} [ \hat{\xi}_\alpha   ,\hat{\varphi}_\beta]  
+ \hat{\varphi}_\alpha  [\hat{a}_{\mu},\hat{\varphi}_\beta]  \Bigr)  \  .
\eeqa

The one-loop approximation amounts to keeping only 
the quadratic terms in $a$ and $\varphi$ in the above expressions,
neglecting the higher-order terms, i.e., the O($a^3$) term and the
O($a^4$) term in $S_{\rm b}$ and the O($a\varphi^2$) term in $S_{\rm f}$.
The quadratic term in $a$ in (\ref{bosonicaction}) has zero modes
due to the fact that the original model (\ref{action})
has the SU($N$) invariance (\ref{SU_N}).
We thus have to ``fix the gauge'' properly.
Following Ref.~\cite{AIKKT}, we choose the ``gauge-fixing'' term
(which is the reduced model counterpart of the Feynman gauge in 
ordinary gauge theory)
and the corresponding Faddeev-Popov ghost term as
\beqa
S_{\rm g.f.} &=& -\frac{1}{2 } \tr ([\hat{x} _\mu  , \hat{a}_\mu]^2)  \ , \n
S_{\rm gh} &=& - \tr ([\hat{x}_\mu , \hat{b}]
[\hat{x}_\mu + \hat{a}_\mu , \hat{c}]) \ .
\label{gaugefixing}
\eeqa
$\hat{a}_\mu$ in the ghost action $S_{\rm gh}$ can be neglected within the
one-loop approximation.
Integration over the ghost fields $\hat{b}$, $\hat{c}$ 
then yields $ \{\Delta (x) \}^2 $,
where $\Delta (x) $ is 
defined as
\beq
\Delta (x) = \prod _{i<j} ( x_{ i } - x_{ j}  )^2  \ .
\eeq

Therefore, the partition function at the one-loop approximation
can be written as
\beq
Z_{\rm IIB} ^{\rm (1-loop)} =\int \dd x  \, \dd a 
\,  \ee ^{-S_{\rm G}} \, \{\Delta (x) \}^2 \, 
Z_{\rm f} ^{\rm (1-loop)} [x,a] \ ,
\label{effectiveaction}
\eeq
where 
\beq
S_{\rm G} =
\sum_{i<j}  ( x_{ i } - x_{ j}  )^2  | a_{\mu ij} | ^2  \  ,
\label{gaussian}
\eeq
and the one-loop approximated fermion integral is defined as
\beqa
Z_{\rm f} ^{\rm (1-loop)} [x,a] &=& \int \dd \xi \, \dd \varphi 
~ \ee ^{-S_{\rm f} ^{(2)}}  \ , \\
S_{\rm f} ^{(2)}&=& 
 - \frac{1}{2} \,
(\tilde{\Gamma} _{\mu})_{\alpha\beta} \tr \Bigl( \hat{\varphi}_\alpha  
[\hat{x}_{\mu},\hat{\varphi}_\beta] 
 - \hat{\varphi}_\alpha    [\hat{\xi}_\beta,\hat{a}_{\mu}]
 - \hat{a}_{\mu} [ \hat{\xi} _\alpha  , \hat{\varphi}_\beta]   \Bigr)  \ .
\eeqa

In what follows we calculate $Z_{\rm f} ^{\rm (1-loop)} [x,a]$ explicitly.
We first note that $S_{\rm f} ^{(2)}$ can be written as
\beq
S_{\rm f} ^{(2)}= 
- \frac{1}{2} \, \sum _{i \neq j}
 \tilde{\varphi}  _{\alpha ji} 
(\tilde{\Gamma} _\mu)_{\alpha\beta} (x_{i \mu} - x_{j \mu} ) 
\tilde{\varphi} _{\beta ij} 
-   \frac{1}{2} \, \sum _{i  j }
\xi_{i \alpha}   M' _{i\alpha , j\beta} \xi _{j\beta}  \ , 
\label{action_quad_ferm}
\eeq
where $\tilde{\varphi} _{\alpha ij}$ 
in (\ref{action_quad_ferm}) are defined by
\beq
\tilde{\varphi } _{\alpha ij} =
\varphi_{\alpha ij} - \frac{(x_{i \rho} - x_{j \rho } ) }
{(x_i - x_j )^2 }
( \tilde{\Gamma} _{\rho} ^ \dag \tilde{\Gamma} _\sigma)_{\alpha\beta} 
a_{\sigma ij} (\xi_{i\beta} - \xi_{j\beta}) \ ,
\eeq
and $M' _{i\alpha , j\beta}$ is a $16N$ $\times$ $16N$ matrix
given as
\beqa
M' _{i \alpha , j\beta}
&=&  \frac{(x_{i\rho} - x_{j\rho})}{(x_i - x_j)^2}
( \tilde{\Gamma} _\mu \tilde{\Gamma} _{\rho}^\dag 
\tilde{\Gamma} _\sigma )_{\alpha\beta}
( a_{\mu ji}  a_{\sigma ij} - a_{\mu ij}  a_{\sigma ji} )
~~~~~ \mbox{for}~~~i \neq j  \ , 
\label{detM}
\\
M' _{i \alpha , i\beta} &=&
- \sum _{j\neq i} M' _{i \alpha , j\beta}  \  .
\label{defMprime}
\eeqa
Integration over $\tilde{\varphi}$ can now be done
yielding $  \{\Delta (x) \}^8 $.
We then integrate out $\xi_{N\alpha}$ using the delta functions
in (\ref{measure_fermion}), yielding
a factor of $1/N^8$ followed by a replacement
\beq
\xi_{N\alpha} \Rightarrow - \sum _{j=1} ^{N-1} \xi_{j\alpha}  
\eeq
in (\ref{action_quad_ferm}).
The integration over the $\xi_{i\alpha}$ ($i=1,\cdots , (N-1)$)
yields $\Pf \, M $.
The matrix $M$ is a $16(N-1)$ $\times$ $16(N-1)$ complex matrix
defined as
\beq
M_{i\alpha , j\beta} 
= M' _{i \alpha , j\beta}  - M' _{N \alpha , j\beta}  
-M' _{i \alpha , N\beta}  + M' _{N \alpha , N\beta}  \ ,
\label{defM}
\eeq
where indices $i$ and $j$ run from 1 to $N-1$.
Note that there are identities
\beq
M'_{j\alpha , i\beta}=M'_{i\alpha , j\beta} ~~~,~~~
M'_{i\beta , j\alpha}=- M'_{i\alpha , j\beta}  \ ,
\eeq
and similarly for $M_{i\alpha , j\beta}$.
This means in particular that $M_{i\alpha , j\beta}$ is an
anti-symmetric matrix, 
regarding each of $(i \alpha)$ and $(j \beta)$ as a single index.
Thus the one-loop approximated
fermion integral is obtained as
\beq
Z_{\rm f} ^{\rm (1-loop)}[x,a] 
=  \{\Delta (x) \}^8 \frac{1}{N^8} \Pf \, M  \ .
\label{one-loopdet}
\eeq
We have checked numerically that indeed
\beq
Z_{\rm f} [A] \simeq Z_{\rm f} ^{\rm (1-loop)}[x,a] ~~~;~~~
(A_\mu)_{ij}  = x_{i \mu} \, \delta _{ij} + a_{\mu ij}  \ ,
\label{approx}
\eeq
holds when the $x_i$'s are well-separated and $a_{\mu ij}$ are
generated with the distribution $\ee ^{-S_{\rm G}}$, where
$S_{\rm G}$ is given by (\ref{gaussian}).
Note that the size of the matrix $M$ in (\ref{one-loopdet})
is of O($N$), whereas the size of the matrix ${\cal M}$ 
in (\ref{fulldet}) is of O($N^2$).
The huge reduction
is essentially because the integration over the fermionic
off-diagonal elements has been done explicitly yielding
$\{\Delta (x) \}^8 $ in (\ref{one-loopdet}).
This is the substantial gain from using the one-loop approximation.


First we note that 
the model (\ref{effectiveaction}) as it stands
has a singularity\footnote{This can be seen more clearly
by rescaling $a_{\mu ij}$ as in (\ref{rescaleBA}).
Then, one finds that all the $x_{i \mu}$-dependence of the 
partition function is contained in the Pfaffian, 
which has the singularity.} 
at $x_{i\mu}=0$ due to the 
singularity in (\ref{detM}).
Therefore if one simulates the model (\ref{effectiveaction}),
the distribution of $x_{i}$ collapses to the origin.
We recall that such an ultraviolet 
singularity does not exist in the original IIB matrix model.
On the other hand, the one-loop approximation is valid only
when $x_{i}$'s are well separated from each other.
Therefore, we have to introduce a UV cutoff
to the distribution of $x_{i}$ in order to make the model meaningful.
As a UV cutoff, we introduce in the action the term given as
\beq
S_{\rm cut}=  \sum_{i < j} f(\sqrt{(x_i - x_j)^2})  \ ,
\label{cutoff}
\eeq
where the function $f(x)$ is taken to be
\beq
f (r) =
\left\{
\begin{array}{ll}
\frac{\kappa }{2 \ell ^2} (r-\ell)^2 & \mbox{for}~~~r < \ell \\
0 & \mbox{for}~~~r \ge \ell  \ .
\end{array}
\right.
\label{def_f}
\eeq
The dimensionless ``spring constant'' $\kappa$ 
should be taken to be large enough
to prevent the $x_i$'s from coming closer to each other
than the cutoff $\ell$ (See Figs.~\ref{fig:EV} and \ref{fig:EV6D}.).
Thus we arrive at
the low-energy effective theory of the IIB matrix model (\ref{action})
\beq
Z_{\rm LEET} = \int \dd x \, \dd a \,  \ee ^{-S_{\rm G}-S_{\rm cut}} 
\,  \{\Delta (x) \}^{10} \, \Pf \, M \  ,
\label{LEETdef}
\eeq
which is written in terms of bosonic variables only.
We have omitted the irrelevant constant factor $\frac{1}{N^8}$
in (\ref{one-loopdet}).

In fact, the UV cutoff parameter $\ell$ can be scaled away
from the theory (\ref{LEETdef})
by rescaling the variables as
$x_{i\mu} \mapsto \ell x_{i\mu}$ and
$a_{\mu ij} \mapsto \frac{1}{\ell}  a_{\mu ij}$.
This means that the dependence of the results on the UV cutoff
$\ell$ is determined completely on dimensional grounds.
In particular, dimensionless quantities are independent of $\ell$.
When we are interested in dimensionful quantities, and in particular
in their large $N$ behavior, we have to know the $N$-dependence of 
the UV cutoff $\ell$.
In Ref.~\cite{AIKKT}, it was argued that
$\ell$ should be taken to be $N$-independent,
based on a reasonable assumption that
the ultraviolet behavior of the space-time structure of 
the IIB matrix model is controlled by the SU(2) matrix model.
The issue is also 
addressed in the four-dimensional version of the IIB matrix
model in Ref.~\cite{AABHN}.
There, an $N$-independent UV cutoff was shown to be generated dynamically 
by treating the full model nonperturbatively 
instead of making perturbative expansions around diagonal matrices.
We therefore take $\ell = 1$ for all $N$ in the following.

The Pfaffian $\Pf \, M$ in 
(\ref{LEETdef})
is complex in general.
This poses the notorious ``sign problem'', when
one tries to study the model by Monte Carlo simulation.
Note, however, that the problem is simply inherited from the
original IIB matrix model (\ref{action}), 
as can be seen from (\ref{approx}).
In the present work, 
we take the absolute value $|\Pf \, M|$ and throw away the phase
by hand.
This corresponds to taking the absolute value 
$|\Pf \, {\cal M}|$ of the Pfaffian in the IIB matrix model
and studying its low-energy effective theory.

To summarize, the model we simulate is given by the partition function
\beq
Z = \int \dd x \, \dd a \,  \ee ^{-S_{\rm G}-S_{\rm cut}} 
\,  \{\Delta (x) \}^{10} \, |\Pf \, M| \  .
\label{OLmodelnew}
\eeq
$S_{\rm G}$ and $S_{\rm cut}$ are given by 
(\ref{gaussian}) and (\ref{cutoff}) respectively,
and $M$ is defined through (\ref{detM}), (\ref{defMprime})
and (\ref{defM}).
Note that the model (\ref{OLmodelnew})
still has the 10D Lorentz invariance.
We would like to investigate whether the 10D Lorentz invariance
breaks down, say, to 4D Lorentz invariance.
The model corresponding to the six-dimensional version
of the IIB matrix model can be obtained similarly.

The validity of
the one-loop approximation for studying the low-energy dynamics
of supersymmetric large $N$ reduced models has been
addressed in Ref.~\cite{AABHN} by studying
the four-dimensional version of the IIB matrix model
without any approximations.
It was found that the large $N$ behavior of the 
extent of the distribution of $x_{i}$ agrees with the 
prediction from the one-loop, low-energy effective theory.
We therefore expect that
a low-energy phenomenon such as 
SSB of Lorentz invariance
can be studied with the low-energy effective theory.

On the other hand we note that from a technical point 
of view the one-loop approximation offers certain advantages
which are essential in the present work.
We recall that even in the four-dimensional version of the IIB matrix
model, the largest $N$ one can achieve for the full model
(without the one-loop approximation) is $N=48$ using supercomputers.
In order to detect even a small trend of SSB
of Lorentz invariance from 10D to 4D, say, one would expect that
$N$ should be larger than $4^4 = 256$.
Due to the fact that the size of the matrix $M$ in (\ref{OLmodelnew})
is order of $N$ smaller than the size of the matrix ${\cal M}$ 
in (\ref{fulldet}),
the one-loop approximation enables us to reach $N=512$ 
with reasonable effort.
Details of the algorithm used in
the Monte Carlo simulations are presented in the Appendix A.
The algorithm is a variant of the Hybrid Monte Carlo algorithm \cite{HMC}
which is one of the standard algorithms used in the study
of systems with dynamical fermions.
The computational effort of the algorithm is estimated to be
O($N^3$)$\sim$O($N^{7/2}$), which should be compared with an
estimate O($N^5$) for the algorithm used for the full model 
in Ref.~\cite{AABHN}.

\vspace*{1cm}

\section{Results for the distribution of $x_{i}$}
\setcounter{equation}{0}
\label{results}

\begin{figure}[hbt]
\begin{center}
\hspace{1cm}
    \includegraphics[height=14cm,angle=270]{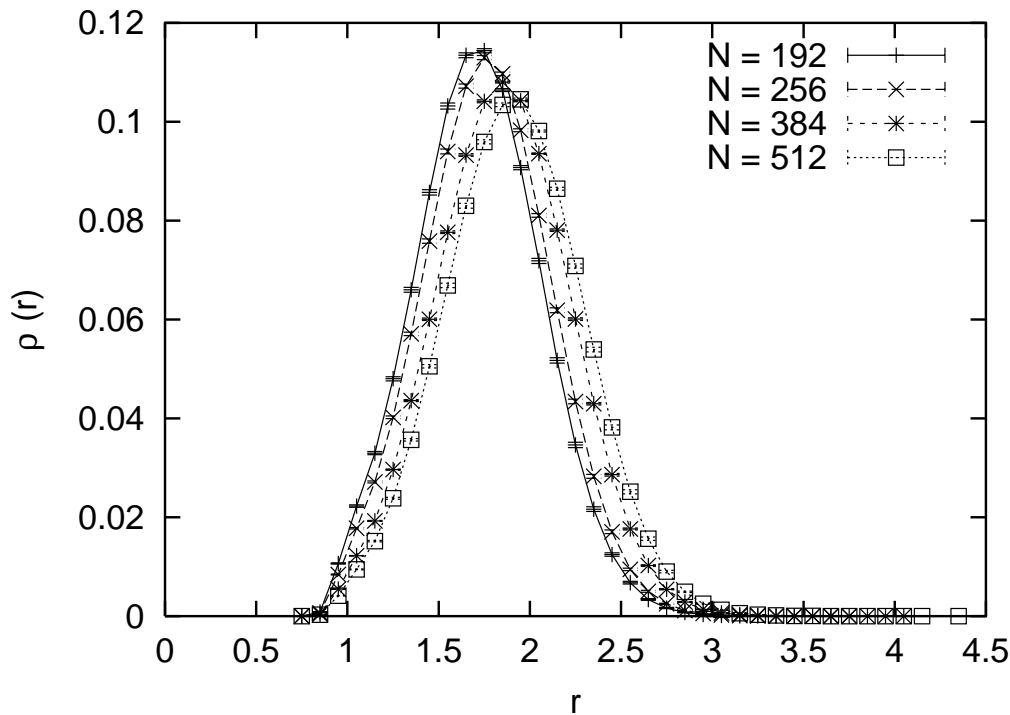}
\end{center}
    \caption{The distribution of distances $\rho(r)$ for $D=10$ is
     plotted against $r$ for $N=192$, 256, 384 and 512.
}
\label{fig:EV}
\end{figure}

\begin{figure}[hbt]
\begin{center}
\hspace{1cm}
    \includegraphics[height=9cm]{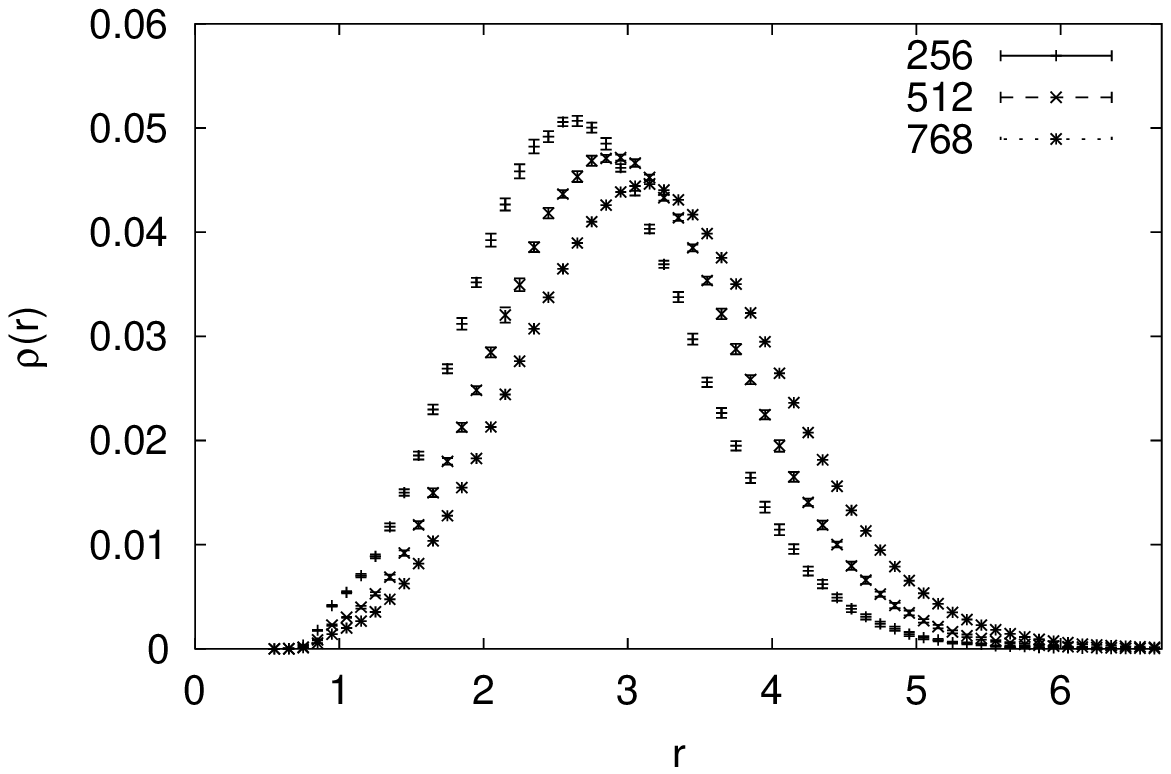}
\end{center}
    \caption{The distribution of distances $\rho(r)$ for $D=6$ is
     plotted against $r$ for $N=256$, 512 and 768.
}
\label{fig:EV6D}
\end{figure}

First we look at the distribution $\rho (r) $
of the distance $r$ between $x_i$'s,
where the distance between two arbitrary points $x_i \neq x_j$
is measured by $\sqrt{(x_i - x_j)^2}$.
In Fig.\ \ref{fig:EV}, we plot the results 
for $D=10$ with $N=192$, 256, 384 and 512.
Fig.\ \ref{fig:EV6D} shows the results 
for $D=6$ with $N=256$, 512 and 768.
We note that in both cases 
the distribution at small $r$ falls off
rapidly below $r\sim \ell$, where $\ell=1$ 
is the UV cutoff introduced
in (\ref{def_f}).
The small penetration 
into the $r < \ell$ region is due to $\kappa$ being finite.
However, the results show that
the values of $\kappa$ we have taken ($\kappa = 300$ for $D=10$ and
$\kappa = 100$ for $D=6$)
are large enough to make the penetration reasonably small.

In order to see the spontaneous breakdown of Lorentz invariance,
we consider the moment of inertia tensor of $N$ points 
$x_{i}$ ($i=1,\cdots ,N$)
in a flat $D$-dimensional space-time\footnote{Such a quantity has 
also been studied in Refs.~\cite{HNT,AIKKT}.} .
It can be defined as
\beq
T _{\mu\nu} = \frac{2}{N(N-1)}  \sum _{i < j} 
(x_{i\mu} - x_{j\mu} ) (x_{i\nu} - x_{j\nu} ) \ ,
\label{defT}
\eeq
which is a $D \times D$ real symmetric matrix.
The $D$ eigenvalues $\lambda _1 > \lambda _2 > \cdots > \lambda _{D} >  0$ 
of the matrix $T$
represent
the principal moments of inertia.
We measure $\lambda _\mu$ for each configuration
and take an average $\langle \lambda _\mu \rangle$
over all the configurations generated
by the Monte Carlo simulation.
If Lorentz invariance is not spontaneously broken,
$\langle \lambda _\mu \rangle$ must be all equal
in the large $N$ limit, representing an isotropic distribution 
of $x_{i}$.
We therefore search for a trend which differs from such a large $N$
behavior.

We first note that if the system that describes the dynamics of 
$x_{i}$ were 
a simple branched polymer in a flat $D$-dimensional space-time, 
spontaneous breakdown of Lorentz invariance would certainly not occur,
and moreover, the large $N$ behavior of $\langle \lambda _\mu \rangle$
could be expected to be $\langle \lambda _\mu \rangle \sim N ^{1/2}$.
This is due to the fact that the Hausdorff dimension 
of a branched polymer is $d_{\rm H} =4 $ 
irrespectively of the dimension $D$ of the space-time in which 
it is embedded.
For this we note that the extent of the distribution of $x_{i}$
is given by 
\beq
R = \sqrt{T_{\mu\mu}} =  \sqrt{\sum _{\mu =1}^D \lambda _{\mu}} \  ,
\label{defR}
\eeq
which is related to the number $N$ of points through
$N \sim (R/\ell) ^{d_{\rm H}}$.
The $\ell$ is the UV cutoff taken to be $\ell =1$.
Therefore, we find $R \sim N ^{1/d_{\rm H}} \sim N^{1/4} $,
which leads to the announced large $N$ behavior
$\langle \lambda _\mu \rangle \sim N ^{1/2}$.

\begin{figure}[htbp]
  \begin{center}
    \includegraphics[angle=270,width=0.8\textwidth]{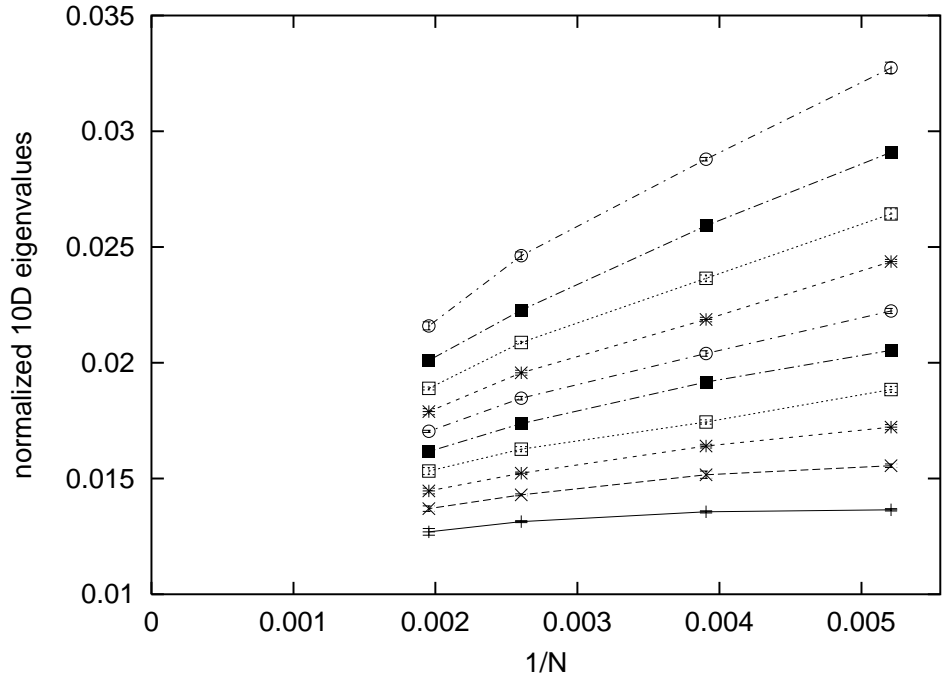}
    \caption{The 
10 eigenvalues of the moment of inertia tensor normalized by
$N^{1/2}$ are plotted against $1/N$ for $N=192,256,384,512$.}
    \label{fig:10Dev}
  \end{center}
\end{figure}

\begin{figure}[htbp]
  \begin{center}
    \includegraphics[angle=270,width=0.8\textwidth]{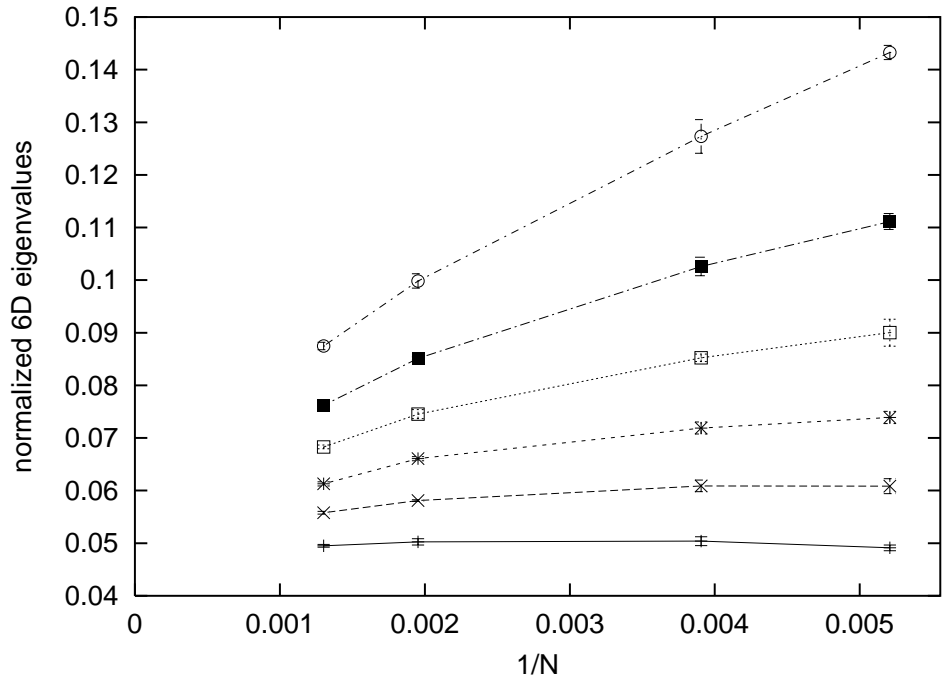}
    \caption{
The 6 eigenvalues of the moment of inertia tensor 
normalized by $N^{1/2}$ are plotted against 
$1/N$ for $N=192,256,512,768$.}
    \label{fig:6Dev}
  \end{center}
\end{figure}

In Fig.~\ref{fig:10Dev} we plot the normalized eigenvalues 
$\langle \lambda _\mu \rangle/N^{1/2}$ against $1/N$
for $D=10$ with $N=192,256,384,512$.
We find that the smallest (normalized) eigenvalue
$\langle \lambda _{10} \rangle/N^{1/2}$ is almost constant
and the larger ones are monotonously decreasing 
towards the same constant.
Thus the observed large $N$ behavior suggests that
there is {\em no} spontaneous breakdown of Lorentz invariance
and the large $N$ behavior of the extent of the $x$-distribution
is consistent with a simple branched-polymer prediction.
In particular, we see no trend for a gap developing between
the fourth and the fifth largest eigenvalues,
which could have been observed if the 10D Lorentz invariance
were broken down to a four-dimensional one.
In Fig.~\ref{fig:6Dev} 
we present the results for $D=6$ with $N=192,256,512,768$.
The qualitative behavior is the same as in $D=10$.

The results for $D=10$ might suggest that the leading
finite $N$ effect is given by $1/N$.
For example, a linear extrapolation in $1/N$
using the data for $N=384,512$ leads to almost the same
results for all the ten (normalized) eigenvalues at $N=\infty$.
It is therefore tempting to speculate that 
finite $N$ effects in the IIB matrix model is given
by a $1/N$ expansion.
Such a form of finite $N$ effects, if it is true,
is in remarkable contrast to the ``bosonic'' case, 
in which the fermions in the IIB matrix model (\ref{action}) 
is omitted by hand.
In that case, the large $N$ behavior of 
correlation functions was determined analytically \cite{HNT}
to all orders in the $1/D$ expansion 
and finite $N$ effects were found to be given by a $1/N^2$ expansion.

For $D=6$, the linear extrapolation gives almost the same results
up to the third smallest eigenvalue, but not for the larger ones.
Thus, as far as reproducing an isotropic distribution
of $x_i$ in the large $N$ limit is concerned,
$D=6$ seems to have larger finite $N$ effects.
This rather counter-intuitive result
might be understood by considering the branched-polymer representation
of the low-energy effective theory \cite{AIKKT}.
The attractive potential between two $x_i$'s 
connected by a bond is given by $r^{-3(D-2)}$, where 
$r =  \sqrt{(x_i - x_j) ^2 }$.
Therefore, it is much stronger for $D=10$ than for $D=6$.
Since the branched-polymer system in both cases is not 
a simple maximal tree
but something more complicated, typically involving a double-tree structure,
it may well be that the stronger the attractive potential is,
the more the distribution of $x_i$ tends to become isotropic.

\vspace*{1cm}

\section{Summary and Discussion}
\setcounter{equation}{0}
\label{summary}

In this paper, we studied the IIB matrix model
(and its six-dimensional version) at large $N$
by using the low-energy effective theory developed 
in Ref.~\cite{AIKKT}.
Unlike in the four-dimensional version studied in Ref.~\cite{AABHN},
the fermion integral yields a Pfaffian (or a determinant in the 6D case)
which is complex in general.
In the present paper we omitted the phase of the Pfaffian by hand in order
to avoid the sign problem in the Monte Carlo simulations,
and we studied the effect of the modulus only.
We have seen that the distribution of $x_i$ becomes more and more
isotropic as we increase $N$.
Based on this observation we conclude that 
if the SSB ever occurs in the original IIB matrix model,
the phase of the Pfaffian induced by fermions
must play a crucial r\^{o}le.

It is interesting to compare with the situation in perturbative 
superstring theory. Also in that case the fermionic degrees 
of freedom can be integrated out explicitly \cite{wiegmann}, leaving
us with an effective bosonic string theory, where the action consists
of three parts: the ordinary bosonic string action, an extrinsic 
curvature term, and (in Euclidean space-time) a purely imaginary
Wess-Zumino-like term. 
This means that if we can neglect the imaginary part, 
the superstring theory is equivalent 
to a bosonic string theory with extrinsic curvature. 
While this equivalence has not been directly disproved, 
it does not have much support, either, from
numerical simulations and analytical calculations 
\cite{extrinsic,polyakov,anaextrin}.

In Ref.~\cite{NV} a deformation of the IIB matrix model
by introducing an integer parameter $\nu$ which couples to
the phase of the Pfaffian
has been considered.
The original IIB matrix model corresponds to $\nu = 1$.
The deformed model continues to be well-defined,
and preserves Lorentz invariance, the SU($N$) symmetry, 
and the cluster property.
In this language the present work corresponds to
the study of the case $\nu = 0$ using the low-energy effective theory.
The opposite extreme, the limit $\nu = \infty$,
has been studied analytically in Ref.~\cite{NV}, 
and the spontaneous breakdown of Lorentz invariance has been discovered.
Of particular interest is 
the fact that different conclusions have been obtained for
$\nu = \infty$ and for $\nu = 0$,
which already implies that a phase transition should occur
in between.
If the original IIB matrix model ($\nu = 1$) belongs to the
same phase as $\nu = \infty$, 
we can investigate the dynamics of the IIB matrix model by studying
the $\nu = \infty$ model
using Monte Carlo simulation as suggested in
Ref.~\cite{NV}.
It would be extremely interesting if we can obtain
a flat four-dimensional space-time in that way.
We hope that the Monte Carlo technique developed in the present work
is useful for such future studies and eventually 
enables us to explore the dynamics of the IIB matrix model.

%



\vspace*{1cm}

\section*{Acknowledgments}

We would like to thank R.J.\ Szabo and G.\ Vernizzi for carefully
reading the manuscript.
J.\ A.\ and K.N.\ A.\ acknowledge the support of MaPhySto,
a center for mathematical physics and stochastics, funded 
by the Danish National Research Foundation.
K.N.\ A.\ acknowledges partial support from a postdoctoral fellowship
from the Greek Institute of National Fellowships (IKY).
J.\ N.\ is supported by the Japan Society for the Promotion of
Science as a Postdoctoral Fellow for Research Abroad. 
The computation has been carried out on 
VPP500 at High Energy Accelerator Research Organization (KEK),
VPP700E at The Institute of Physical and Chemical Research (RIKEN),
SR8000 at Computer Centre, University of Tokyo,
and SX4 at Research Center for Nuclear Physics (RCNP) of Osaka
University.
J.\ N.\ is grateful to T.~Ishikawa, S.~Ohta, A.I.~Sanda, H.~Toki,
S.~Uehara and Y.~Watanabe for many kind helps in using the supercomputers.
This work is supported by the Supercomputer Project (No.99-53)
of KEK.

\newpage


\section*{Appendix A: The algorithm for the simulation}
\setcounter{equation}{0}
\renewcommand{\theequation}{A.\arabic{equation}}
\hspace*{\parindent}


In this appendix, we explain the algorithm\footnote{Ref.\ \cite{Weingarten} 
gives an overview of effective algorithms for dynamical fermions.}
we use
for the Monte Carlo simulation of our model defined by
(\ref{OLmodelnew}).

We first simplify the model by introducing
rescaled variables $b_{\mu ij}$ as
\beq
b_{\mu ij} = \sqrt{(x_i - x_j)^2 } \, a_{\mu ij}  \  .
\label{rescaleBA}
\eeq
The model we have to simulate then becomes
\beq
Z = \int \dd x \, \dd b 
\,  \ee ^{-S_{\rm G}-S_{\rm cut}} \,  |\Pf \, M | \  ,
\label{OLmodel}
\eeq
where
\beq
S_{\rm G} =
\sum_{i<j}  | b_{\mu ij} | ^2  \ ,
\label{Sgb}
\eeq
and (\ref{detM}) can be rewritten in terms of $b_{ij}$ as
\beq
M' _{i \alpha , j\beta}
=  
\frac{(x_{i\rho} - x_{j\rho})}{\{(x_i - x_j)^2 + (\epsilon \ell)^2 \}^2}
(\tilde{\Gamma} _\mu \tilde{\Gamma} _{\rho} ^\dag 
\tilde{\Gamma} _\sigma )_{\alpha\beta}
(b_{\mu ji}  b_{\sigma ij} - b_{\mu ij}  b_{\sigma ji} )
~~~~~ \mbox{for}~~~i \neq j   \ ,
\label{Mprimeb_reg}
\eeq
where we have also introduced
the dimensionless regularization parameter $\epsilon$
to reduce numerical instabilities.
The regularization parameter $\epsilon$ should be taken to be
small enough not to affect the system.
For both, 6D and 10D, cases, we took $\epsilon = 0.1$.
We note also that with this regularization,
the model is well-defined without introducing the cutoff term (\ref{cutoff}).
However, by simulating such a model,
we find that all the $x_i$'s become densely packed within the 
extent $\epsilon \ell$ as we increase $N$.
Therefore, we still need the cutoff term (\ref{cutoff}) in order
to make our model valid as a low-energy effective theory
of the original IIB matrix model.
For convenience we combine
our dynamical variables $x_{i \mu}$ and $b_{\mu ij}$ 
($i\ne j$, $b_{\mu ji}=b_{\mu ij}^*$)
into a Hermitian matrix $B_{\mu}$ as
$(B_\mu)_{ij} = x_{i\mu} \, \delta_{ij} + b_{\mu ij}$.
%
%

The algorithm we use to simulate the model (\ref{OLmodel})
is a variant of the Hybrid Monte Carlo algorithm \cite{HMC}.
The first step of the Hybrid Monte Carlo algorithm is
to apply molecular dynamics \cite{molecule}.
We introduce a conjugate momentum for $B_{\mu ij} $ as
$X_{\mu ij}$, which satisfies 
$X_{\mu ji} = (X_{\mu ij} )^* $; $X_\mu$ are Hermitian
matrices.
The partition function can be rewritten as\footnote{Unlike
in the standard Hybrid Monte Carlo algorithm \cite{HMC}, 
we do not introduce the so-called pseudo-fermions.
Note, in this regard, that our system (\ref{OLmodel})
is different from ordinary field theories with
dynamical fermions in the following respects.
1) The number of fermion flavors should be strictly one
in order to respect supersymmetry.
2) The size of the matrix $\, M \, $ is much smaller than the 
system size.
3) The matrix $\, M \, $ is not sparse.}
\beq
Z = \int \dd X  \, \dd B \,  \ee ^{-H} \ ,
\eeq
where $H$ is the ``Hamiltonian'' defined by
\beq
H = \frac{1}{2} \sum _{\mu i j} | X_{\mu ij} | ^2 
+ S_{\rm G} [B]+ S_{\rm cut} [B] -   \frac{1}{2}\ln | \det M |  \ .
\label{defH}
\eeq
The update of $X_{\mu ij}$ can be done by just
generating $X_{\mu ij} $ with the probability distribution
$\exp (-\frac{1}{2}\sum |X_{\mu  ij}|^2)$.
In order to update $B_{\mu ij}$, we solve the Hamilton equation
given by
\beqa
\frac{\dd B_{\mu ij} (\tau)}{\dd \tau}
&=&  \frac{\del H}{\del X_{\mu ij} }  = X_{\mu  ji}  \\
\frac{\dd X_{\mu ij} (\tau)}{\dd \tau}
&=& - \frac{\del H}{\del B_{\mu ij}  }  
= \frac{1}{4} \Tr \left(
 \frac{\del D }{\del B_{ \mu ij } } D^{-1}  \right)
- \frac{\del S_{\rm G}}{\del B_{\mu  ij} }
- \frac{\del S_{\rm cut}}{\del B_{\mu  ij} }   \ ,
\label{Hamiltonianeq}
\eeqa
where we have defined $D = M^\dag M$.
Note that when taking the derivatives in (\ref{Hamiltonianeq}),
$(B_\mu)_{ij}$ and $(B_\mu)_{ji}$
should be treated as independent complex variables.
Along the ``classical trajectory'' given by 
the Hamilton equation,

(i) $H$ is invariant, 

(ii) the  motion is reversible,

(iii) the phase-volume is preserved; {\it i.e.}
\beq
\frac{\del (B(\tau),X(\tau))}{
\del (B(0),X(0))} = 1  \  ,
\eeq
where $ (B(\tau),X(\tau))$ is the point on the 
trajectory after evolution for fixed $\tau$ 
from $(B(0),X(0))$.
Therefore, generating a new sets of $(B,X)$ 
by solving the Hamilton equation for a fixed ``time'' interval 
$\tau$ satisfies the detailed balance.
This procedure, together with the proceeding
generation of $X_{\mu ij}$  with the Gaussian distribution,
is called ``one trajectory'', which corresponds
to ``one sweep'' in ordinary Monte Carlo simulations.


When solving the Hamilton equation numerically,
we have to discretize the equation.
A discretization which preserves the properties (ii) and (iii)
is known.
The property (i) cannot be preserved and yields a small violation
of the Hamiltonian conservation.
In order to still satisfy the detailed balance exactly,
we can perform a Metropolis accept/reject as the end of each trajectory.

We introduce a short-hand notation for discretized
$X_{\mu} (\tau)$ and $B_{\mu} (\tau)$ as
\beq
X_{\mu} ^{(r)} = X_{\mu}  (r \Delta \tau) ~~~;~~~~~
B_{\mu} ^{(s)} = B_{ \mu} ( s \Delta \tau )  \ .
\eeq
Given the configuration $(B_{ \mu }  ^{(0)},X_{\mu }  ^{(0)})$,
the configuration $(B_{\mu }  ^{(\nu)}, X_{\mu }  ^{(\nu)})$
after the evolution for a fixed time $\tau = \nu \, \Delta \tau $
is obtained by solving the discretized Hamilton equation
\beqa
\label{evolB}
B_{\mu ij} ^{ (\frac{1}{2})} &=& 
B_{\mu ij} ^{(0)}  + \frac{\Delta \tau }{2}
X_{\mu ji} ^{(0)}  \n
B_{\mu ij} ^{(n+\frac{1}{2})} &=& 
B_{\mu ij} ^{(n-\frac{1}{2})} + \Delta \tau 
X_{\mu ji} ^{(n)}  \n
B_{\mu ij}  ^{(\nu)} &=& 
B_{\mu ij} ^{(\nu-\frac{1}{2})} + \frac{\Delta \tau }{2}
X_{\mu ji} ^{(\nu)}  \\
X_{\mu ij}  ^{ (m+1)} &=& X_{\mu ij}  ^{(m)} 
- \Delta \tau \frac{\del H }{\del B_{\mu ij}  } 
( B_\mu ^{(m+\frac{1}{2})} )  \ , 
\label{evolX}
\eeqa
where $n=1,2,\cdots, \nu -1$ and $m=0,1,\cdots, \nu -1$.
Note that the first and the final steps in the evolution (\ref{evolB})
of $B_\mu$ are treated with special care.
This particular discretization, which is called as
``leap-frog discretization'',
preserves the properties (ii) and (iii).
At the end of the trajectory, we make a Metropolis 
accept/reject with the probability
$P=\min (1, \ee ^{- \Delta H})$, where
$\Delta H$ is the difference of the Hamiltonian $H$
defined in (\ref{defH}) for the configurations
$(B_{ \mu }  ^{(0)}, X_{\mu }  ^{(0)})$ and
$(B_{ \mu }  ^{(\nu)}, X_{\mu }  ^{(\nu)})$.
Optimization of $\Delta \tau$ and $\nu$ is discussed in Appendix B.

In the evolution (\ref{evolX}) of $X_\mu$, one needs to calculate
\beq
\Tr \left(   \frac{\del D }
{\del B_{\mu ij}   } D^{-1}  \right)  
= \Tr \left(   \frac{\del M }
{\del B_{\mu ij}  } M^{-1} \right)  +
 \Tr \left(   \frac{\del M ^\dag}
{\del B_{\mu ij}  }  M^{\dag -1} \right)  \ .
\eeq
If we write the first term as
$T_{\mu ij}$, the second term can be written
as $T_{\mu ji} ^*$. 
In what follows, we calculate $T_{\mu ij}$ explicitly.
We first note that
\beq
T_{\mu ij} =
\sum_{k,l=1}^{N-1}\sum_{\alpha \beta}
\frac{\del M_{k\alpha , l\beta}}{\del B_{\mu i j}}
C_{l\beta , k \alpha}
=
\sum_{k,l=1}^{N}\sum_{\alpha \beta}
\frac{\del M '_{k\alpha , l\beta}}{\del B_{\mu i j}}
C'_{l\beta , k \alpha}  \ ,
\eeq
where $C_{i\alpha , j \beta}$ is defined as
\beq
C_{i\alpha , j \beta} = (M^{-1})_{i \alpha , j \beta} \ ,
\label{dominantpart}
\eeq
and $C_{i\alpha , j \beta} '$ is defined as
\beqa
C_{i \alpha , j \beta} ' = C_{i \alpha , j \beta}  ~~~;~~~
C_{i \alpha , N \beta} ' =  - \sum _{k=1}^{N-1} C_{i \alpha , k \beta} \n
C_{N \alpha , j \beta} ' =  - \sum _{k=1}^{N-1} C_{k \alpha , j \beta} ~~~;~~~
C_{N \alpha , N \beta} ' = \sum _{k,l=1}^{N-1} C_{k \alpha , l \beta} \  .
\label{Cprime}
\eeqa
The indices $i,j$ in (\ref{dominantpart}) and (\ref{Cprime})
run from 1 to $N-1$.
We further rewrite $T_{\mu ij}$ as
\beq
T_{\mu ij} =
\sum_{k = 1}^{N}
\sum_{l \ne k}^{N} 
\sum_{\alpha \beta}
\frac{\del M '_{k\alpha , l\beta}}{\del B_{\mu i j}}
 C''_{l\beta , k \alpha}  \ , 
\eeq
where we have introduced $C_{i \alpha , j \beta} ''$ through
\beq
C_{i \alpha , j \beta} '' = C_{i \alpha , j \beta}  '
                        - C_{j \alpha , j \beta}  '  \ .
\eeq
Then we calculate 
$\frac{\del M '_{k\alpha , l\beta}}{\del B_{\mu i j}}$
explicitly, which yields
\beq
T_{\mu ij} =
\sum_{\alpha \beta}
\frac{(x_{i\rho} - x_{j \rho})}{\{(x_i - x_j)^2 + (\epsilon \ell) ^2\}^2}
\{ (\tilde{\Gamma} _ \sigma \tilde{\Gamma} _ {\rho} ^\dag  
\tilde{\Gamma} _ \mu)_{\alpha \beta}
- (\tilde{\Gamma} _ \mu \tilde{\Gamma} _ {\rho} ^ \dag  
\tilde{\Gamma} _ \sigma)_{\alpha \beta} \}
b_{\sigma ji} C_{i \beta , j \alpha} '''
\eeq
for $i\ne j$ and
\beqa
T_{\mu ii} &=&
\sum_{k \ne i}^{N} \sum_{\alpha \beta}
\left[
\frac{\delta _{\mu \rho} }{\{(x_i - x_k)^2+(\epsilon \ell)^2\}^2}
- \frac{4 (x_{i \mu} - x_{k \mu})(x_{i \rho} - x_{k \rho})}
{\{(x_i - x_k)^2+ (\epsilon \ell)^2\}^3 }  \right] \n
&&~~~~~
(\tilde{\Gamma} _ \sigma \tilde{\Gamma} _ {\rho} ^\dag 
\tilde{\Gamma} _ \tau)_{\alpha \beta}
(b_{\sigma k i}b_{\tau  i k} - b_{\sigma i k} b_{\tau k i} )
C_{i \beta , k \alpha} '''  \ ,
\eeqa
where we defined $C_{i \alpha , j \beta} '''$ through
\beqa
C_{i \alpha , j \beta} ''' &=& C_{i \alpha , j \beta}  ''
                         + C_{j \alpha , i \beta}  '' \n
                         &=&  C_{i \alpha , j \beta}  '
                        - C_{j \alpha , j \beta}  '
                        - C_{i \alpha , i \beta}  '
                        + C_{j \alpha , i \beta}  ' \ .
\eeqa

Let us comment on the required computational effort
of our algorithm.
The dominant part comes from calculating the inverse in 
(\ref{dominantpart}), which requires $O(n^3)$ arithmetic operations,
where $n$ is the size of the matrix to be inverted.
In the present case $n$ is of O($N$).
In order to keep the acceptance rate at
the Metropolis accept/reject procedure reasonably high,
one has to decrease the step size $\Delta \tau$ as one goes to larger $N$.
We have seen from simulations that the step size should be taken to be
$\Delta \tau \sim \frac{1}{\sqrt{N}}$, which is consistent with
a general formula $\Delta \tau \sim V ^{-1/4}$ in Ref.~\cite{GIKP}, 
where the system size $V$ should be
replaced by O($N^2$) in our case.
Accordingly, the number of steps for one trajectory increases
as $ \nu \sim \sqrt{N}$.
Therefore, the required computational effort is estimated to be 
$O(N^{7/2})$.

In fact, one can reduce the computational effort
by omitting the Metropolis accept/reject
procedure, since in that case one can keep the step size $\Delta \tau$ 
fixed to a small constant for all $N$.
The required computational effort becomes $O(N^{3})$.
The price one has to pay is that the algorithm then suffers from
a systematic error due to the small violation of the Hamiltonian
conservation.
The systematic error can be estimated to be O($(\Delta \tau)^2$),
see Appendix B.

\vspace*{1cm}


\section*{Appendix B: Optimization of the algorithm}
\setcounter{equation}{0}
\renewcommand{\theequation}{B.\arabic{equation}}
\hspace*{\parindent}

In this appendix, we first 
discuss the optimization of the parameters
in the algorithm 
with the Metropolis accept/reject procedure.
We then move on to the case
when the Metropolis accept/reject procedure is omitted.
The algorithm and the parameters used for each run
is also described.

To start with,
one can actually generalize the algorithm described in Appendix A
by taking the step size $\Delta\tau _x$ for the
diagonal elements to be different\footnote{This is 
not the case for the full model \cite{AABHN}, where
$\Delta\tau$ should be taken universally for each element of
$A_{\mu ij}$ in order to respect the SU($N$) invariance.
In the present case, since the SU($N$) ``gauge'' invariance is 
fixed by the ``gauge-fixing'' (\ref{gaugefixing}),
we only have to respect the invariance under permutation of the
SU($N$) indices.} from the step size $\Delta\tau$ for the
off-diagonal elements of $B_{\mu ij}$ and $X_{\mu ij}$.
We have fixed the ratio $\Delta \tau_x / \Delta\tau$ to be
the ratio of the mean magnitudes of
$x_{i\mu}$ and $b_{\mu ij}$,
which we found from simulations to be approximately
$1$ for $D=6$ and $1/2$ for $D=10$, respectively.

We still have two parameters in the algorithm, i.e.,
$\Delta\tau$ and $\nu$ (the number of molecular dynamics steps for
each trajectory),
which can be optimized in a standard way \cite{GKS,hybridR}.
The key point of the optimization is that
the autocorrelation time (in units of accepted trajectory)
depends only on  $\nu  \, \Delta\tau = \tau$, but not on  
$\Delta\tau$ and $\nu$ separately.
This allows us to perform the optimization in two steps.
First, one fixes $\nu \, \Delta\tau  = \tau$
and optimize $\Delta\tau$ so that the effective speed
of motion in the configuration space, given by acceptance rate
times $\Delta\tau$, is maximized.
Second, using the optimized $\Delta\tau$ for each $\tau$,
one minimizes a typical autocorrelation time in
units of molecular dynamics step with respect to $\tau$.

\begin{figure}[htbp]
  \begin{center}
    \includegraphics[width=0.65\textwidth]{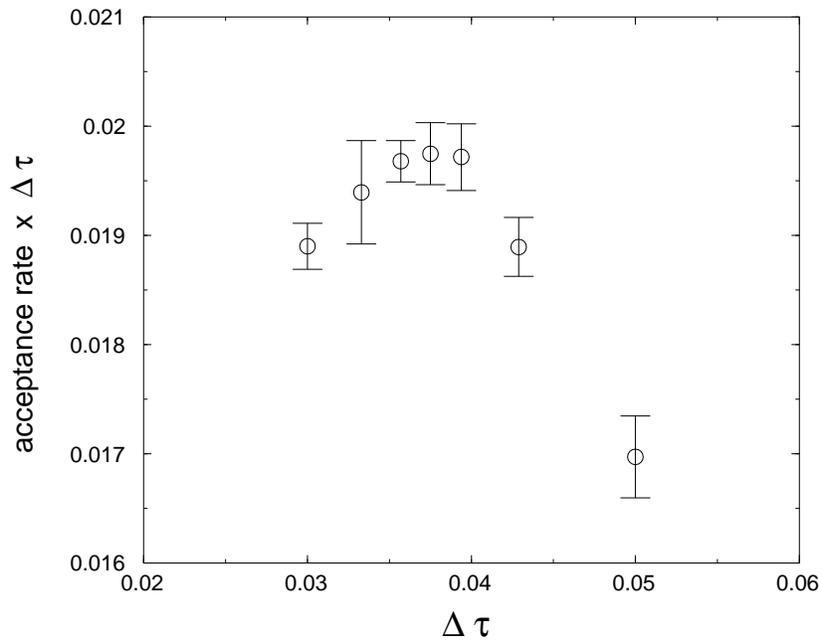}
    \caption{The acceptance rate times $\Delta\tau$  
is plotted against $\Delta\tau$ for $D=6,
      N=16$ with fixed $\tau = 1.5$.}
    \label{fig:H6DN16}
  \end{center}
\end{figure}

In Fig. \ref{fig:H6DN16}, we show the acceptance rate times 
$\Delta\tau$  as a
function of $\Delta\tau$ for $D=6$, $N=16$
with fixed $\tau = 1.5$.
We find that the optimal $\Delta\tau$  is $0.0375$.
For $D=6, N=32$ with fixed $\tau = 1.5 $, 
we find that the optimal 
$\Delta\tau$ is $0.02142$, which is smaller than for $N=16$ as expected.
For both cases, the acceptance rate at the optimal $\Delta\tau$ 
is found to be $50 \sim 60 \%$.

\begin{figure}[htbp]
  \begin{center}
    \includegraphics[width=0.65\textwidth]{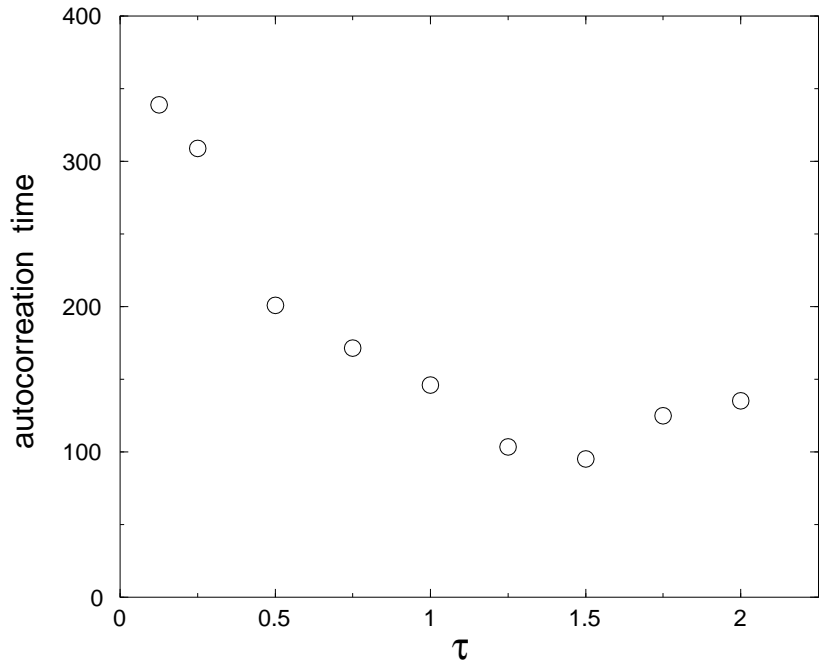}
    \caption{The autocorrelation time (in units of molecular dynamics
step) of the extent $R$ of the $x_{i}$-distribution
is plotted against $\tau$ for $D=6,N=16$.
$\Delta \tau$ is chosen for each $\tau$ so that the acceptance rate
times $\Delta \tau$ is maximized.}
    \label{fig:H6DN16a}
  \end{center}
\end{figure}

Using the optimal $\Delta \tau$ obtained in the above way
for each $\tau$,
we minimize a typical autocorrelation time 
(in units of molecular dynamics step)
with respect to $\tau$.
Here, we measure the autocorrelation time
of the extent $R$ of the $x_{i}$-distribution defined in (\ref{defR})
and plot it as a function of $\tau$.
Fig. \ref{fig:H6DN16a} shows the result for $D=6$ with $N=16$.
We see that it has a minimum around $\tau \sim 1.5$.
Similar experiments for $N=32$ showed that
the optimal $\tau$ is almost independent of $N$.

\begin{figure}[htbp]
  \begin{center}
    \includegraphics[width=0.6\textwidth]{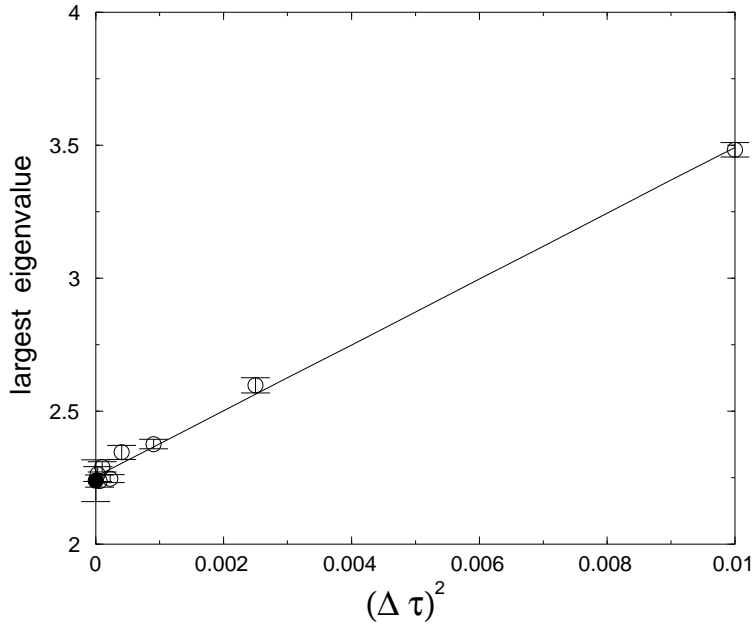}
    \caption{The expectation value of the largest eigenvalue 
of the moment of inertia tensor is measured by simulations
{\em without} Metropolis accept/reject procedure.
The result is plotted against $(\Delta\tau)^2$ for
$D=6, N=16$ with fixed $\tau=1.5$.
The filled circle at $\Delta\tau=0$ represents the result
obtained by a simulation {\em with} Metropolis accept/reject procedure.
   The straight line is a fit to the predicted $(\Delta\tau)^2$
behavior of the systematic error.}
    \label{fig:e6DN16}
  \end{center}
\end{figure}

When we omit the Metropolis accept/reject procedure,
the algorithm suffers from a systematic error, as is explained
in Appendix A.
Here we fix $\nu \, \Delta\tau$
to the optimal $\tau$
obtained for the case with the Metropolis accept/reject procedure
and study the $\Delta\tau$ dependence of the systematic error.
As a quantity which shows a large systematic error,
we take $\langle \lambda _1 \rangle$,
the expectation value of the largest eigenvalue of 
the moment of inertia tensor defined by (\ref{defT}).
In Fig. \ref{fig:e6DN16}, 
we plot the result against $(\Delta\tau)^2$ 
for $D=6, N=16$ with $\tau=1.5$.
The systematic error is seen to vanish as O($(\Delta\tau)^2$),
which can be also understood theoretically by using the analysis
described in Ref.~\cite{hybridR}.

Finally, let us comment on the algorithms and the parameters
we used in our simulations at large $N$.
The runs for $D=6$ with $N=192,256$
were made by the algorithm including 
the Metropolis accept/reject procedure.
The parameters are 
$\Delta \tau = 0.0048, \nu = 60$ for $N=192$,
and $\Delta \tau = 0.004, \nu = 50$ for $N=256$.
The numbers of configurations used for the measurements
are 560 and 1732 for $N=192$ and $N=256$, respectively.
(The optimization described in this Appendix was not completed
when we started these runs. Accordingly, we needed 
significantly larger numbers of configurations compared to
the cases below.)
For the other cases, we omitted the 
Metropolis accept/reject procedure
in order to obtain a sufficient statistics.
The parameters are 
$\Delta\tau = 0.0075, \nu = 200$ for $D=10$ with $N=192,256,384,512$,
and $\Delta\tau = 0.005, \nu = 300$ for $D=6$ with $N=512,768$.
The numbers of configurations used for the measurements are
88, 40, 84, 44 for $D=10$ with $N=192, 256, 384, 512$, 
and 64, 50 for $D=6$ with $N=512, 768$, respectively.



\end{document}